\documentclass[preprint,12pt]{elsarticle}

\usepackage{amssymb}
\usepackage[version=4]{mhchem}
\usepackage{graphics}
\usepackage{caption}
\usepackage{subcaption}
\usepackage{graphicx}
\usepackage{wrapfig}
\usepackage{lscape}
\usepackage{rotating}
\usepackage{booktabs}
\biboptions{sort&compress}

\begin{document}

\begin{frontmatter}


\title{High-temperature oxidation and nitridation of substoichiometric zirconium carbide in isothermal air}

\author[inst1,inst2]{Matthew T. Konnik}

\affiliation[inst1]{organization={Center for Hypersonics and Entry Systems Studies},
            addressline={University of Illinois at Urbana-Champaign}, 
            city={Urbana},
            state={Illinois},
            postcode={61801}, 
            country={USA}}
            
\author[inst1,inst4]{Trey Oldham}
\author[inst1,inst2]{Allison Rzepka}
\author[inst1,inst2]{Vincent Le Maout}
\author[inst1,inst2]{Kelly A. Stephani}
\author[inst1,inst4]{Francesco Panerai}

\affiliation[inst2]{organization={Department of Mechanical Science and Engineering},
            addressline={University of Illinois at Urbana-Champaign}, 
            city={Urbana},
            state={Illinois},
            postcode={61801}, 
            country={USA}}

\affiliation[inst4]{organization={Department of Aerospace Engineering},
            addressline={University of Illinois at Urbana-Champaign}, 
            city={Urbana},
            state={Illinois},
            postcode={61801}, 
            country={USA}}

\begin{abstract}
The influence of nitrogen on the oxidation behavior of hot-pressed zirconium carbide was investigated using a flow-tube furnace at temperatures ranging from 1000 to 1600 $^{\circ}$C. Mass gain, oxide formation characteristics, and oxide transitions were evaluated at various experimental conditions. Differences in oxidation behavior across the range of temperatures investigated show both kinetic and microstructural dependence with implications pointing to this materials efficacy in ultra-high temperature applications. Results suggest that at temperatures above 1400 $^{\circ}$C, although oxidation mechanisms remain dominant, nitridation and reduction mechanisms may be appreciable enough to require consideration. Supporting discussions regarding polymorphism and microstructural influences are outlined.
\end{abstract}




\end{frontmatter}



Ultra-high temperature (UHT) materials have garnered research interest in applications spanning hypersonics, nuclear engineering, and manufacturing due to their unique ability to withstand harsh environments. Zirconium carbide (ZrC$_x$), a transition metal carbide, is a promising UHT material owing to the excellent thermal, mechanical, and chemical properties it possess \cite{Katoh2013,Wang2008,Mackie1989,Caccia2018,opeka_mechanical_1999}. Despite these advantages, reports on the material response of ZrC$_x$ at higher temperatures are limited, often resulting in the usage of better understood transition metal carbides and borides \cite{Fahrenholtz2017}. In particular, the thermochemical response of UHT materials in nitrogen and oxygen, being the main constituents of air, are some of the most important behavioral characteristics to understand. Yet, this has not been thoroughly characterized for ZrC$_x$ at high temperatures.

Numerous breakthrough studies on the response of ZrC$_x$ to oxygen have been performed in the last century and have chronologically investigated higher temperatures. Kuriakose \textit{et al}. investigated the kinetics of ZrC$_x$ oxidation in a heated flow chamber, but were limited to the range of 554-652 °C due to the mechanical stability of the oxide layer formed \cite{Kuriakose1964}. Shimada \textit{et al}. investigated both isothermal and non-isothermal furnace oxidation of ZrC$_x$ powders from 380 to 550 °C with oxygen partial pressures ranging from 1.3 to 7.9 kPa under 39.5 kPa total static pressure \cite{Shimada1990}. Rama Rao \textit{et al}. explored ZrC$_x$ oxidation at temperatures of up to 800 °C \cite{Rao1994} and more recently, Gasparrini \textit{et al}. studied oxidation kinetics in a laboratory furnace at temperatures of up to 1100 °C \cite{Gasparrini2018}. Few studies have been performed at elevated temperatures (up to 1800 °C), yet these were performed in an O$_{2}$ environment at very low pressures \cite{Shabalin}. Wuchina and Opeka have also reviewed the oxidation properties of ZrC$_x$ \cite{fahrenholtz_group_2014}. Each study has observed some aspect of a two-phase oxidation mechanism through oxycarbide formation and carbon precipitation subsequently leading to ZrO$_{2}$ formation. Though, numerous differences exist regarding reported kinetics, as well as resulting microstructure and morphologies of the scale, likely due to differences in substrate stoichiometry, impurity concentration and test environment. 

Nitridation of ZrC$_x$ is much less studied, yet the effects can be drastic. Kieffer \textit{et al} investigated the nitridation behavior of transition metal carbides in an autoclave furnace from 1400 to 1800°C at 1 to 300 atm, reporting carbide to nitride conversions of up to 80 mol\% \cite{Kieffer1970}. A few other studies have also been performed in nitrogen atmospheres up to 2500 °C, all with similar findings regarding the transition of carbide to carbonitride and nitride phases \cite{Klimashin1972,Vilk1968,Neshpor1967,Binder1995}.

While there has been work on both the oxidation and nitridation of ZrC$_x$, there is little to no discussion on how one can influence another. In order to understand the role nitrogen exposure can have on the oxidation of ZrC$_x$, as well as provide some of the first oxidation data available beyond 1100°C at atmospheric pressure, this work focuses on furnace experiments in which a ZrC$_x$ substrate is pre-treated in nitrogen before air oxidation.


ZrC$_x$ sputtering targets (99.9\%, MSE Supplies) were used as received to create test specimens sized 12.5 $\times$ 6 $\times$ 6 mm. Note the target was fabricated using hot isostatic pressing at 2273 K. Specimens were sectioned from the bulk material using a low speed precision cutter (IsoMet) attached with a diamond sectioning blade (IsoMet 20 LC, Buehler). Verification of the ZrC$_x$ system, described in Table ,was confirmed as explained in Supplementary Note I of the supplementary information (SI).

\begin{table}[h]
\centering
\begin{tabular}{c c c}
\toprule
C:Zr Ratio & Purity & Porosity\\ 
\midrule \hline
0.64 & 99.9\% & 12\% \\
\midrule \hline
\end{tabular}
\caption{As-recieved ZrC$_x$ parameters}
\label{ZrC_descrip}
\end{table}

Oxidation experiments were performed in a horizontal tube furnace (CM 1730, CM Furnaces Inc.). 
Test specimens were then placed in a high-purity alumina crucible and inserted into the furnace. The furnace was then ramped to temperature (at a rate of 200 $^{\circ}$C/hr up to 800 $^{\circ}$C and 100 $^{\circ}$C/hr thereafter) while flowing 2 lpm of nitrogen (measured using a variable area flow meter). Upon reaching the desired temperature (1000 to 1600 $^{\circ}$C), gas composition was changed to a mixture of 20\% oxygen and 80\% nitrogen (0.4 lpm O$_{2}$, 1.6 lpm N$_{2}$) for the desired exposure duration (10 to 40 minutes). Upon completion of the desired oxidation exposure time, gas flow was switched back to nitrogen and the furnace was ramped down to room temperature with an identical ramp profile. Samples mass and dimensions were recorded prior to and after furnace exposure. 

Samples were then analyzed using scanning electron microscopy (SEM, Quanta FEG 450 ESEM, FEI Company), x-ray photoelectron spectroscopy (XPS, Kratos Axis ULTRA, Kratos Analytical), and x-ray diffraction (Cu K-$\alpha$ source; D8 Advance; Bruker Corperation) according to details described in Note II of the SI.

\begin{figure}
    \centering
    \begin{subfigure}[b]{0.49\textwidth}
        \centering
        \includegraphics[width=0.95\textwidth]{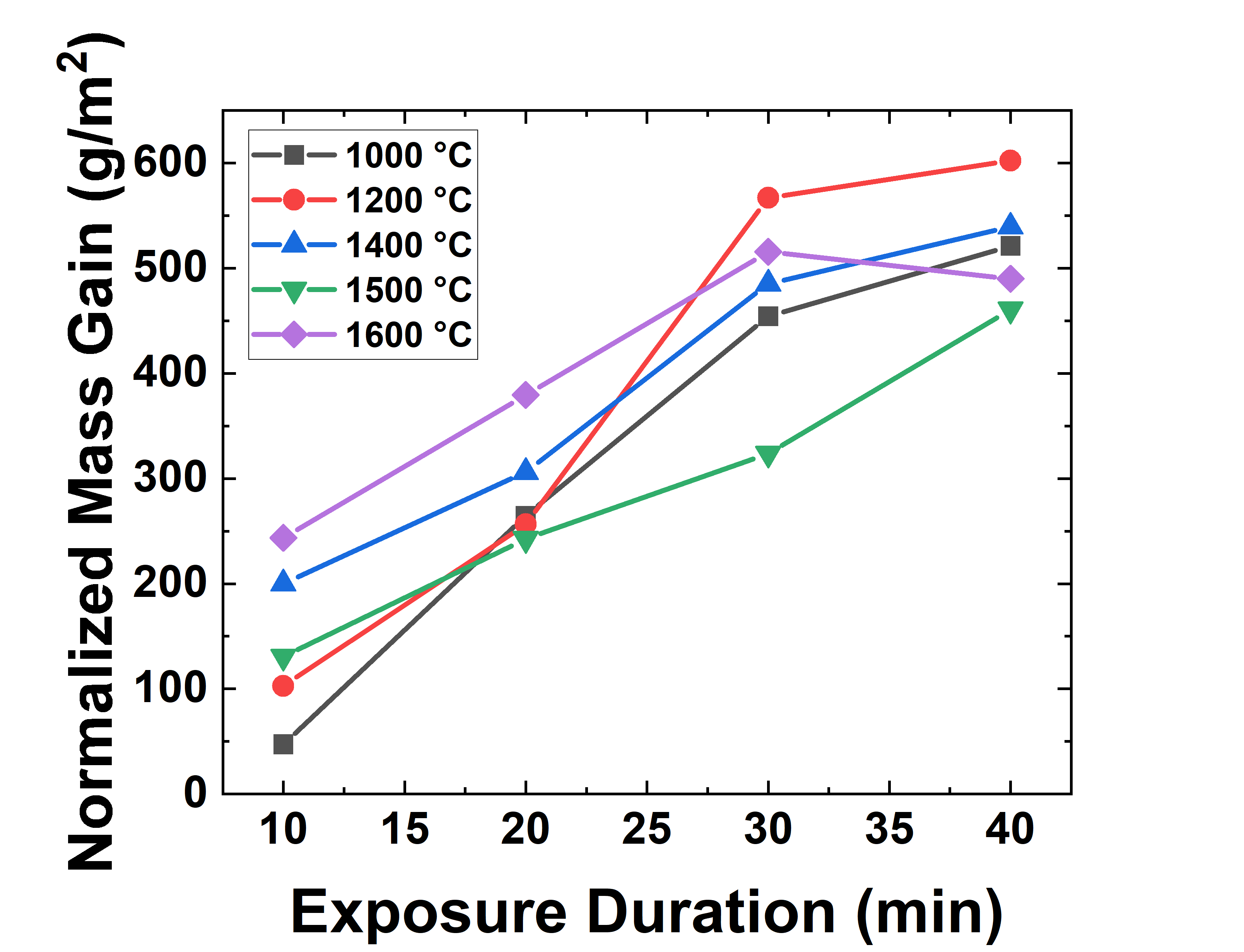}
        \caption{Normalized mass gain plotted versus exposure duration for specified oxidation temperatures.}
        \label{massvtime}
    \end{subfigure}
    \hfill
    \begin{subfigure}[b]{0.49\textwidth}
        \centering
        \includegraphics[width=0.95\textwidth]{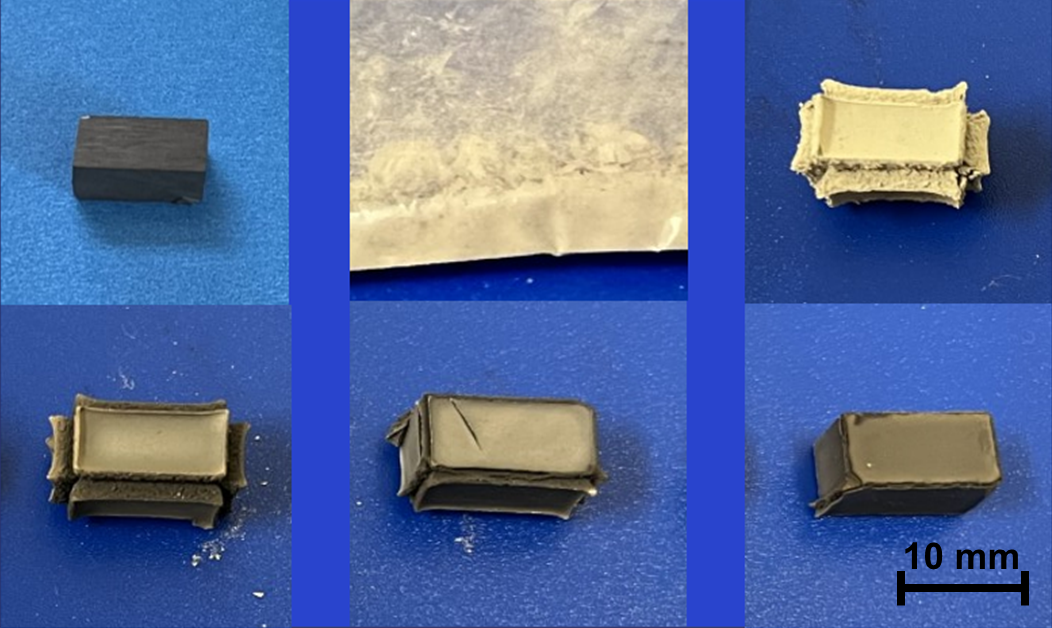}
        \caption{Top row, left to right: unoxidized specimen, 1000 $^{\circ}$C, 1200 $^{\circ}$C. Bottom row, left to right: 1400 $^{\circ}$C, 1500 $^{\circ}$C, 1600 $^{\circ}$C. All oxidized specimens exposed for 40 minutes.}
        \label{specimens}
    \end{subfigure}
    \caption{Mass gain versus time profile and examples of test specimens.}
    \label{fig1}
\end{figure}

Figure \ref{massvtime} shows total mass gain, as a relative percentage, versus air exposure duration, a test matrix is available in Section III of the SI. The 1000 $^{\circ}$C 10 minute condition was run in triplet and the standard deviation of normalized mass gain found to be 11.64 g m$^{-2}$. Varying the exposure duration of ZrC$_x$ specimens in high temperature oxidizing environments revealed cross-over in relative mass gain across the time series, suggesting a deviation from a pure Arrhenius rate from. As exposure duration increases, the lower temperature regime begins to dominate in total mass increase. Specifically, specimens oxidized at 1200 $^{\circ}$C overtake higher oxidation temperature mass gains for exposure durations longer than 20 minutes. We find that kinetic, morphological, and microstructural deviations across temperatures are likely all major contributors to the observed mass gain behavior.

The set of competing reactions occurring during ZrC$_x$ oxidation are given by \cite{Gasparrini2018}:

\begin{equation}
\centering
\label{eq1}
\ce{ZrC + \frac{1}{2}(1-x)O_{2}_{(g)} -> ZrC_{x}O_{1-x} + (1-x)C} 
\end{equation}

\begin{equation}
\centering
\label{eq2}
\ce{ZrC_{x}O_{1-x} + \frac{1}{2}(1 + 3x)O_{2}_{(g)} -> ZrO_{2} + (x)CO_{2}_{(g)}} 
\end{equation}

\begin{equation}
\centering
\label{eq3}
\ce{C + O_{2}_{(g)} -> CO_{2}_{(g)}}
\end{equation}

Reaction (\ref{eq1}) describes the conversion at the reaction front and interfacial region, capable of producing stable zirconium oxicarbides, zirconia, and free carbon. Reaction (\ref{eq2}) describes the production of zirconia from any oxycarbide formations, which is accompanied by carbon dioxide outgassing. Reaction (\ref{eq3}) describes a separate mechanism for which free carbon formations present in the substrate or intermediate layer can form gaseous carbon dioxide independently of zirconia formation. The reaction in Eq. \ref{eq1} describes a reaction that only produces solid state products, unlike Eqs. \ref{eq2} and \ref{eq3} lead to gaseous by-products while resulting in a net mass gain or mass loss, respectively. The consequence is that a small deviation in kinetic rate dependence with temperature between these three reaction mechanisms can lead to large deviations in mass gain behavior. 

Any nitridation occurring prior to oxygen exposure introduces further complexity by altering the activation energy and structure associated with the aforementioned oxidation mechanisms. Preliminary furnace experiments performed under pure nitrogen provided evidence of a direct nitridation mechanism, such as that described in Eq. \ref{eq4}, active at the temperatures in this study. Given that furnace ramp rate is constant, each oxidation temperature explored is associated with a different nitrogen exposure duration, which combined with varying temperatures leads to dissimilar degrees of nitridation in the substrate. When considering a substoichiometric ZrC$_x$, increasing degrees of nitrogen incorperation into carbon vacancy cites through intermediate pathways of Eq. \ref{eq4} is likely to increase the oxidation resistance of the substrate, contributing to the suppressed mass gain of higher temperatures observed in Fig. \ref{massvtime}.

\begin{equation}
\centering
\label{eq4}
\ce{2ZrC + N_{2}_{(g)} -> 2ZrN + 2C}
\end{equation}

\begin{figure}
    \centering
    \includegraphics[width=0.98\textwidth]{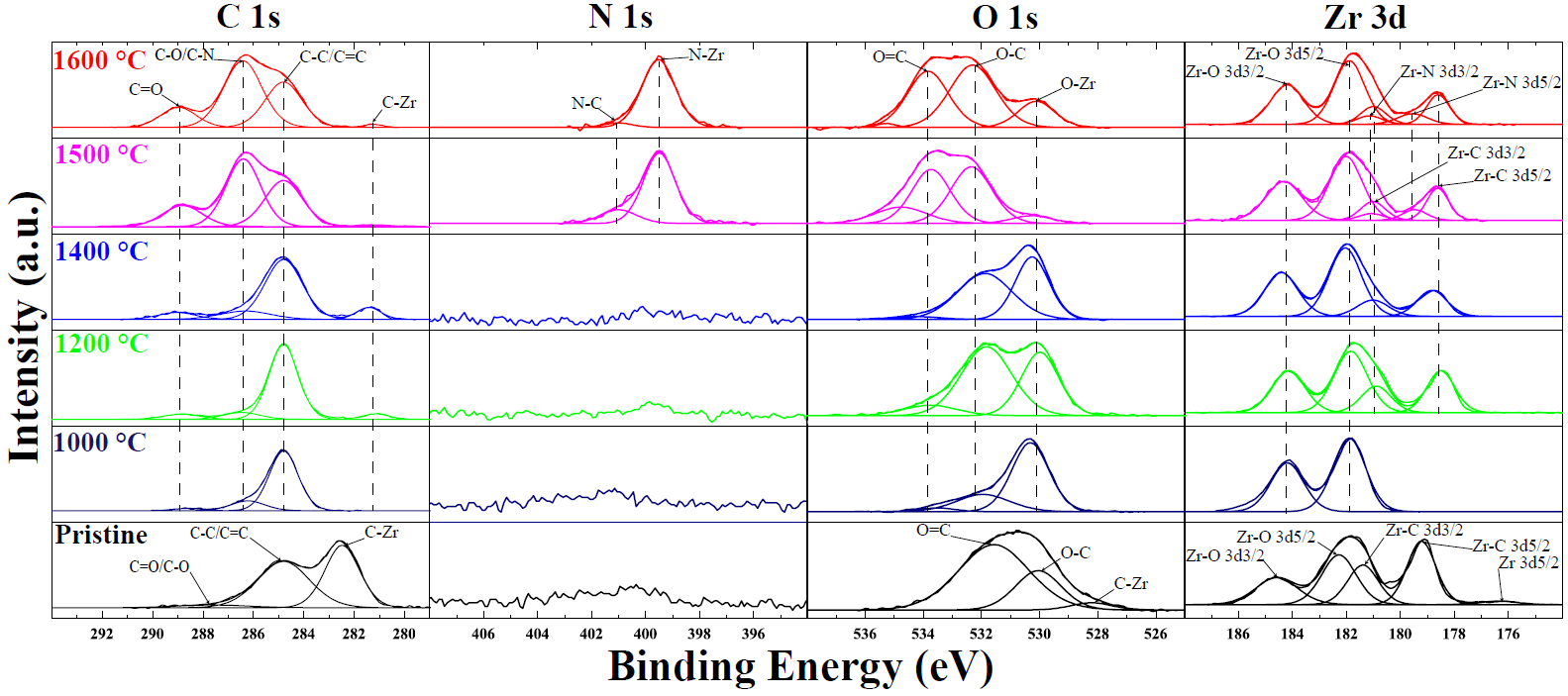}
    \caption{X-ray photoelectron spectroscopy high resolution scans and fits of the C1s, N1s, O1s, and Zr3d orbitals. Each row is associated with a specimen treated for 20 min (if applicable) at conditions designated on the left.}
    \label{XPS}
\end{figure}

Fig. \ref{XPS} shows high resolution XPS scans for each specimen treated in air for 20 minutes at increasing temperatures. Where applicable, the spot was focused on a cross section of the sample encompassing the reaction front and surroundings (see SI Note IV for details). Cross sections were prepared for XPS by first sectioning with an IsoMet 20 LC diamond blade, followed by SiC polishing to 1200 grit, and lapped with diamond films down to 3 $\mu$m. Specimens were cleaned in an ultrasonic bath of isopropanol and outgassed in between preparation steps. For the 1000 $^{\circ}$C oxidation case, the collected powder was analyzed as the scale had no mechanical integrity. The majority of XPS signal was attributed to Zr-C, Zr-O, C-C, and typical adventitious carbon associations \cite{long_phase_2014,hakansson_high-resolution_1993,hou_mechanisms_2020,majumdar_xray_1991,sinha_xps_1987,kumar_xps_1988,sarma_xpes_1980,morgan_comments_2021,hauser_hybrid_2019,gu_promotion_2019,liu_exploration_2017}. 
For samples exposed to higher than 1400 $^{\circ}$C, analysis showed an uptake in N1s signal and a corresponding pertubation in the Zr3d signal around 179.5eV, consistent with the presence of a metallic Zr-N type bonding \cite{ul-hamid_effect_2020,badrinarayanan_xps_1989,calderon_v_chemical_2015} . Drastic increases in the carbonyl and C-O/C-N groups were also present in this temperature regime, which requires further investigation.

Observation of Zr-N type bonding above 1400 $^{\circ}$C suggests that the ZrC$_x$ substrate may have been sufficiently nitridated during furnace ramp up such that not all nitrogen-containing compounds were consumed during oxidation. Though, the combined carbothermic-reduction nitridation (Eq. \ref{eq5}) mechanism often seen in the synthesis of ZrN using a ZrO$_{2}$ precursor \cite{fu_synthesis_2004,zhao_synthesis_2019} remains plausible given the production of free carbon during ZrO$_{2}$ formation (Eq. \ref{eq1}) \cite{Katoh2013,Shimada1990,Rao1994,david_tem_2013} and the dominant presence in the C1s signal in the XPS spectra shown in Fig. \ref{XPS}. It is crucial to note that the oxidation of ZrC$_x$ at temperatures exceeding 1400  $^{\circ}$C approaches the minimum temperatures used in the synthesis of ZrC$_x$ and ZrN, which typically leverage a carbothermic reduction or combined reduction-nitridation mechanism of ZrO$_{2}$ \cite{sun_study_2018,sondhi_theoretical_2013,harrison_processing_2016}. While synthesis of these products is usually performed by homogenizing ZrO$_{2}$ and carbon graphite powders and heating in an inert environment, an oxidizing environment can produce the same conditions locally at the ZrC$_x$-ZrO$_{2}$ interface where solid state and gasous carbon products are created during oxidation or direct nitridation. While the XPS results provide compelling evidence of nitridation mechanisms at these temperatures, the decrease in the overall mass gain of furnace specimens from 30 and 40 minutes at 1600 $^{\circ}$C is further indication of active reduction mechanisms. As such, during air exposure of ZrC$_x$ above 1400  $^{\circ}$C on relevant timescales, the process described in Eqs. 1-3, can be accompanied by nitridation and carbothermal reduction mechanisms, as presented in synthesis literature  \cite{zhao_synthesis_2019,parkison_deconvolution_2016}, described in Eqs. \ref{eq5} - \ref{eq6}:

\begin{equation}
\centering
\label{eq5}
\ce{2ZrO_{2} + N_{2}_{(g)} + 4C -> 2ZrN + 4CO_{(g)}}
\end{equation}

\begin{equation}
\centering
\label{eq6}
\ce{ZrO_{2} + 3C -> ZrC + 2CO_{(g)}}
\end{equation}

\begin{figure}[ht]
    \centering
    \includegraphics[width=0.7\textwidth]{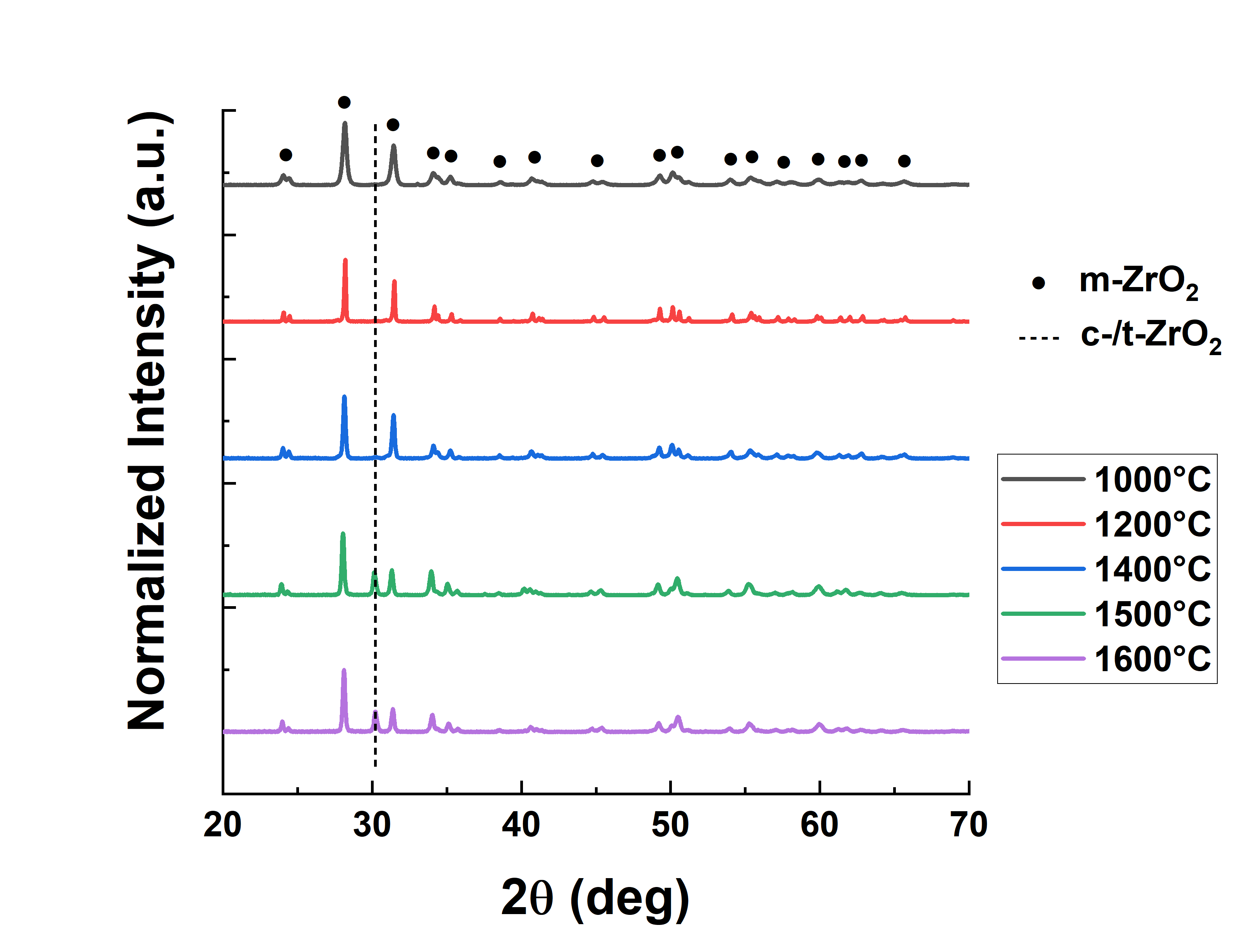}
    \caption{X-ray diffraction profiles of the surface of oxide formations associated with each temperature investigated. Profiles shown were extracted from 30 minute exposure test specimens.}
    \label{XRD}
\end{figure}

These reaction mechanisms and their intermediate steps, compete with the oxidation processes in Eqs. \ref{eq1} - \ref{eq3} to describe the deviations in mass gain dominance in temperature and time observed Fig \ref{massvtime}. While slowed down by the processes described in \ref{eq4} - \ref{eq6}, mass gain via oxidation remains predominant across the entire temperature range investigated, as clearly shown by the ZrO$_{2}$ signal in the XPS scans, and also in the XRD measurements shown later in Fig. \ref{XRD}. Reduction mechanisms may manifest themselves in mass gain data as a response to long exposures where reaction front oxygen flux drops off in time due to oxide growth.

The kinetics discussed thus far are further influenced by the evolving morphology of the oxide. Zirconia equilibrium phases are well known; monoclinic zirconia (m-ZrO$_{2}$) is stable at and above room temperature, and that tetragonal zirconia (t-ZrO$_{2}$) is stable at temperatures between roughly 1173 $^{\circ}$C and 2360 $^{\circ}$C on heating and 987 $^{\circ}$C and 2360 $^{\circ}$C on cooling at atmospheric pressure \cite{Gotsch2016,Gurak2018,Block1985}. Although the temperatures for pre-existing zirconia morphologies are relatively well defined, preferential zirconia phase nucleation with temperature during ZrC$_x$ oxidation are less understood. Figure \ref{specimens}, shows specimen oxide formations across a range of temperatures exposed for 40 minutes. Drastic differences in both the appearance and mechanical integrity of the oxide layer formation exist in temperature. Both these features can be attributed to the zirconia morphology developed during oxidation, phase transition (if any) and accompanying Martensistic transformation, which introduces high mechanical integrity by way of residual stress formation in the oxide \cite{Jin2005}, on cooling. A discussion on the mechanical integrity of oxide formations is presented in Supplementary Note V of the SI.

While the temperatures explored in this study are too high for many x-ray diffractometers to observe nucleation of ZrO$_{2}$ in-situ, a footprint of the previous morphology can still be found, allowing for a more definitive metric of evaluation. Whereas the majority of the oxide formation undergoes a  phase-transformation to m-ZrO$_{2}$ on cooling, some stabilized t-ZrO$_{2}$ crystallite formations which nucleated during oxidation at temperature remain after cooling. Figure \ref{XRD} shows this phenomenon by way of XRD analyses performed at room temperature, post-oxidation on the surface of the oxide scale. For the specimens oxidized at a sufficiently high temperature (1400 - 1600 $^{\circ}$C), these tetragonal crystallites are visible at a $2\theta$ angle of 30.26 degrees, suggesting the presence of a stabilizing agent, such as nitrogen, integrated in the t-ZrO$_{2}$ lattice \cite{lerch_nitridation_2005,cheng_nitrogen-containing_1991,cheng_role_1993}. In contrast, for specimens oxidized below the nucleation temperature required to form a sufficient quantity of t-ZrO$_{2}$ (1000 and 1200 $^{\circ}$C), there is no signature present. These observations were found to be independent of exposure duration. 

It is unknown how drastically morphological differences across temperature alter rate coefficients, but differences in formation energy and lattice configuration, transformation toughening (when oxidizing at temperatures fluctuating around phase transition temperatures), local density differences in the oxide due to variation in morphology, and increased crystallite dislocations and defects due to polymorphic zirconia formation can all play a part. While these morphological aspects can influence the effective chemical kinetics, microstructural dependencies on temperature can also be coupled to these phenomena. 

\begin{figure}[ht]
    \centering
    \begin{subfigure}[b]{0.24\textwidth}
        \centering
        \includegraphics[width=0.95\textwidth]{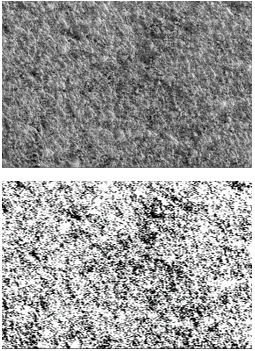}
        \caption{1200 $^{\circ}$C}
        \label{1200_pore}
    \end{subfigure}
    \hfill
    \begin{subfigure}[b]{0.24\textwidth}
        \centering
        \includegraphics[width=0.95\textwidth]{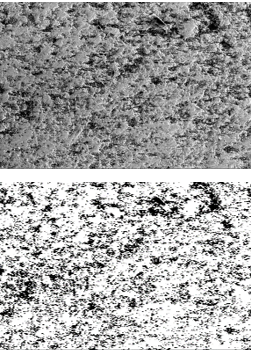}
        \caption{1400 $^{\circ}$C}
        \label{1400_pore}
    \end{subfigure}
    \hfill
    \begin{subfigure}[b]{0.24\textwidth}
        \centering
        \includegraphics[width=0.95\textwidth]{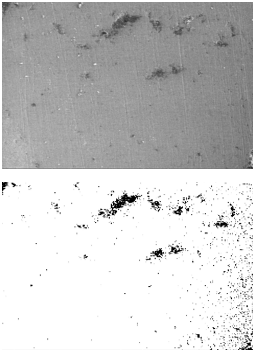}
        \caption{1500 $^{\circ}$C}
        \label{1500_pore}
    \end{subfigure}
    \hfill
    \begin{subfigure}[b]{0.24\textwidth}
        \centering
        \includegraphics[width=0.95\textwidth]{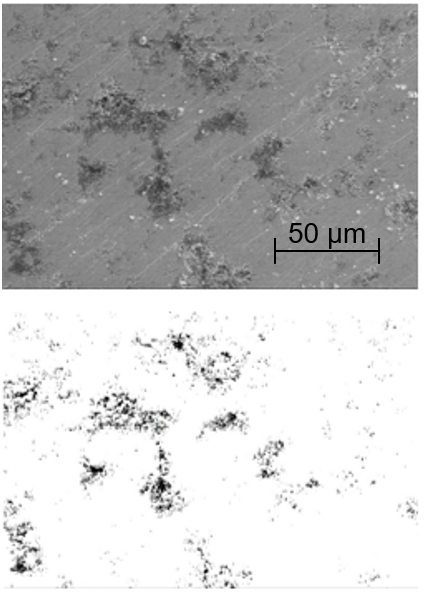}
        \caption{1600 $^{\circ}$C}
        \label{1600_pore}
    \end{subfigure}
    \caption{Example micrographs and corresponding binary images of oxide formations from sectioned and polished specimens exposed to oxygen for 20 minutes at designated temperatures.}
    \label{porosity}
\end{figure}

Figure \ref{porosity} shows example micrographs of pore formations in oxide formations (after sectioning and polishing of sufficiently high mechanical integrity specimens) across temperatures with corresponding binary images formed from thresholding. Binary images were then used to form approximations for resulting porosity in the field of view (assuming isotropic pore formations). Resulting porosity approximations are averages over at least three 200 $\mu$m field of view images per exposure duration, and over at least two exposure durations per temperature. Given this approach, porosities in zirconia formations created at 1200, 1400, 1500, and 1600 $^{\circ}$C are determined to be 35\%, 18\%, 2.5\%, and 2\%, respectively. 

Aside from the morphological aspects, the porosity of an oxide formation can be impacted by grain growth size at a given temperature. Grain boundary mobility is influenced by and proportional to temperature, meaning that larger crystallites will form at higher oxidation temperatures \cite{humphrey}. With larger grains at higher temperatures, comes a smaller total grain boundary area. Then at higher temperatures, a reactants diffusion into the material, and outflow of gaseous oxidation by-products have less opportunity to initiate pore formation. In contrast, at lower temperatures with smaller crystal formations, total grain boundary area is relatively high (by way of a larger number of formations), giving a higher probability for gaseous products to form pores at temperature. This method of void formation in the oxide is, of course, tied to the kinetics by way of carbon outgassing rate and differential concentration of oxygen. In turn, these quantities are dependent on porosity and tortuosity of the oxide formation.

In this study, we have investigated the high temperature oxidation response of zirconium carbide specimens in the presence of a constant nitrogen flow, using a flow-tube furnace environment. Relative mass gain versus exposure duration profiles were obtained for temperatures ranging from 1000 to 1600 $^{\circ}$C and the kinetic, morphological, and microstructural influences were discussed. Nitridation and reduction mechanisms were found to be influential in the material response, especially at at sufficiently high temperatures, suppressing the expected rate of mass gain above 1200 $^{\circ}$C. These processes are suggested to take place primarily in the scale-substrate interfacial region, suggested by the presence of a nitride phase in cross section analyses, but not in external scale investigations. Phase transitions and polymorphism affecting the specimens at the studied temperatures also present an influence on the quality of oxide formed, as well as potential influence on mass gain through differing formation energies and defect modes. Finally, porosity differences in oxide formations were found to be as high as 32\%, with a decreasing trend in porosity with increasing temperature, creating a dense, potentially passivating oxide formation (depending on specimen geometry) at temperatures in excess of 1400 $^{\circ}$C. Observations presented in this work act as the only current investigation regarding the influence of nitrogen on the bulk oxidation of zirconium carbide at ultra-high temperatures. This work further serves also as a basis for future investigations across the regimes which influence aggregate thermochemical response behavior of zirconium carbides and potential nitrogen treatments to alter oxidation properties and scale quality.

\section{Acknowledgements}
The work presented in this article was supported by Lockheed Martin Cooperation under grant LMC \#S20-005. The authors acknowledge the use of facilities and instrumentation at the Materials Research Laboratory Central Research Facilities, University of Illinois, partially supported by NSF through the University of Illinois Materials Research Science and Engineering Center DMR-1720633. The authors would also like to thank Dan Hecht from Lockheed Martin for his guidance and many fruitful discussions.

\appendix

 \bibliographystyle{elsarticle-num} 
 \bibliography{refs}

\begin{thebibliography}{10}
\expandafter\ifx\csname url\endcsname\relax
  \def\url#1{\texttt{#1}}\fi
\expandafter\ifx\csname urlprefix\endcsname\relax\def\urlprefix{URL }\fi
\expandafter\ifx\csname href\endcsname\relax
  \def\href#1#2{#2} \def\path#1{#1}\fi

\bibitem{Katoh2013}
Y.~Katoh, G.~Vasudevamurthy, T.~Nozawa, L.~L. Snead, Properties of zirconium
  carbide for nuclear fuel applications, Journal of Nuclear Materials 441
  (2013) 718--742.
\newblock \href {https://doi.org/10.1016/j.jnucmat.2013.05.037}
  {\path{doi:10.1016/j.jnucmat.2013.05.037}}.

\bibitem{Wang2008}
Y.~Wang, Q.~Liu, J.~Liu, L.~Zhang, L.~Cheng, Deposition mechanism for chemical
  vapor deposition of zirconium carbide coatings, Journal of the American
  Ceramic Society 91 (2008) 1249--1252.
\newblock \href {https://doi.org/10.1111/j.1551-2916.2007.02253.x}
  {\path{doi:10.1111/j.1551-2916.2007.02253.x}}.

\bibitem{Mackie1989}
W.~A. Mackie, P.~R. Davis, Single-crystal zirconium carbide as a
  high-temperature thermionic cathode material, IEEE Transactions on Electron
  Devices 36 (1989) 220--224.
\newblock \href {https://doi.org/10.1109/16.21209}
  {\path{doi:10.1109/16.21209}}.

\bibitem{Caccia2018}
M.~Caccia, M.~Tabandeh-Khorshid, G.~Itskos, A.~R. Strayer, A.~S. Caldwell,
  S.~Pidaparti, S.~Singnisai, A.~D. Rohskopf, A.~M. Schroeder, D.~Jarrahbashi,
  T.~Kang, S.~Sahoo, N.~R. Kadasala, A.~Marquez-Rossy, M.~H. Anderson,
  E.~Lara-Curzio, D.~Ranjan, A.~Henry, K.~H. Sandhage, Ceramic–metal
  composites for heat exchangers in concentrated solar power plants, Nature 562
  (2018) 406--409.
\newblock \href {https://doi.org/10.1038/s41586-018-0593-1}
  {\path{doi:10.1038/s41586-018-0593-1}}.

\bibitem{opeka_mechanical_1999}
M.~M. Opeka, I.~G. Talmy, E.~J. Wuchina, J.~A. Zaykoski, S.~J. Causey,
  Mechanical, {Thermal}, and {Oxidation} {Properties} of {Refractory} {Hafnium}
  and zirconium {Compounds}, Journal of the European Ceramic Society 19~(13-14)
  (1999) 2405--2414.
\newblock \href {https://doi.org/10.1016/S0955-2219(99)00129-6}
  {\path{doi:10.1016/S0955-2219(99)00129-6}}.

\bibitem{Fahrenholtz2017}
W.~G. Fahrenholtz, G.~E. Hilmas, Ultra-high temperature ceramics: Materials for
  extreme environments, Scripta Materialia 129 (2017) 94--99.
\newblock \href {https://doi.org/10.1016/j.scriptamat.2016.10.018}
  {\path{doi:10.1016/j.scriptamat.2016.10.018}}.

\bibitem{Kuriakose1964}
A.~K. Kuriakose, J.~L. Margrave, The oxidation kinetics of zirconium diboride
  and zirconium carbide at high temperatures, Journal of the Electrochemical
  Society 111 (1964) 827--831.

\bibitem{Shimada1990}
S.~Shimada, Oxidation kinetics of zirconium carbide at relatively low
  temperatures, Journal of the American Ceramics Society 73 (1990).

\bibitem{Rao1994}
G.~A.~R. Rao, V.~Venugopal, Kinetics and mechanism of the oxidation of zrc,
  Journal of Alloys and Compounds 206 (1994) 237--242.

\bibitem{Gasparrini2018}
C.~Gasparrini, R.~J. Chater, D.~Horlait, L.~Vandeperre, W.~E. Lee, Zirconium
  carbide oxidation: Kinetics and oxygen diffusion through the intermediate
  layer, Journal of the American Ceramic Society 101 (2018) 2638--2652.
\newblock \href {https://doi.org/10.1111/jace.15479}
  {\path{doi:10.1111/jace.15479}}.

\bibitem{Shabalin}
I.~L. Shabalin, Zirconium monocarbide, in: Ultra-High Temperature Materials II,
  Springer Nature Singapore Pte Ltd., Singapore, 2019, pp. 562--567.
\newblock \href {https://doi.org/10.1007/978-94-024-1302-1}
  {\path{doi:10.1007/978-94-024-1302-1}}.

\bibitem{fahrenholtz_group_2014}
E.~J. Wuchina, M.~Opeka, The {Group} {IV} {Carbides} and {Nitrides}, in: W.~G.
  Fahrenholtz, E.~J. Wuchina, W.~E. Lee, Y.~Zhou (Eds.), Ultra-{High}
  {Temperature} {Ceramics}, John Wiley \& Sons, Inc, Hoboken, NJ, 2014, pp.
  361--390.
\newblock \href {https://doi.org/10.1002/9781118700853.ch14}
  {\path{doi:10.1002/9781118700853.ch14}}.

\bibitem{Kieffer1970}
R.~Kieffer, H.~Nowotny, P.~Ettmayer, M.~Fruedhofmeier, {\"U}ber die
  best{\"a}ndigkeit von {\"u}bergangsmetallcarbiden gegen stickstoff bis zu 300
  at, Monatshefte f{\"u}r Chemie 101 (1970) 62--82.

\bibitem{Klimashin1972}
G.~Klimashin, A.~Avgustinik, G.~Smirnov, Karbonitridnye i oksikarbidnye fazy
  titana i tsirkoniya, SSSR Neorg Mater 8~(5) (1972) 843--845.

\bibitem{Vilk1968}
Y.~Vilk, I.~Danisia, Y.~Inekchenko, T.~Ryzhkova, Variations of certain
  physico-chemical properties in the system zirconium carbide - zirconium
  nitride, Russ J Appl Chem 41~(4) (1968) 873--881.

\bibitem{Neshpor1967}
V.~Neshpor, Y.~Vilk, I.~Danisia, Change in the electro- and thermophysical
  properties in pseudobinary alloys along the ray {ZrC$_{0.92}$}-{ZrN$_{0.85}$}
  of the zirconium-nitrogen-carbon system, Powder Metall Met Ceram 6~(1) (1967)
  68--71.

\bibitem{Binder1995}
S.~Binder, W.~Lengauer, P.~Ettmayer, J.~Bauer, J.~Debuigne, M.~Bohn, Phase
  equilibra in the systems {Ti - C - N}, {Zr - C - N} and {Hf - C - N}, J
  Alloys Compd 217~(1) (1995) 128--136.

\bibitem{long_phase_2014}
Y.~Long, A.~Javed, J.~Chen, Z.-k. Chen, X.~Xiong, Phase composition,
  microstructure and mechanical properties of {ZrC} coatings produced by
  chemical vapor deposition, Ceramics International 40~(1, Part A) (2014)
  707--713.
\newblock \href {https://doi.org/10.1016/j.ceramint.2013.06.059}
  {\path{doi:10.1016/j.ceramint.2013.06.059}}.

\bibitem{hakansson_high-resolution_1993}
K.~L. Håkansson, H.~I.~P. Johansson, L.~I. Johansson, High-resolution
  core-level study of {ZrC}(100) and its reaction with oxygen, Physical Review
  B 48~(4) (1993) 2623--2626.
\newblock \href {https://doi.org/10.1103/PhysRevB.48.2623}
  {\path{doi:10.1103/PhysRevB.48.2623}}.

\bibitem{hou_mechanisms_2020}
L.~Hou, Q.~Liang, F.~Wang, Mechanisms that control the adsorption–desorption
  behavior of phosphate on magnetite nanoparticles: the role of particle size
  and surface chemistry characteristics, RSC Advances 10~(4) (2020) 2378--2388.
\newblock \href {https://doi.org/10.1039/C9RA08517C}
  {\path{doi:10.1039/C9RA08517C}}.

\bibitem{majumdar_xray_1991}
D.~Majumdar, D.~Chatterjee, X‐ray photoelectron spectroscopic studies on
  yttria, zirconia, and yttria‐stabilized zirconia, Journal of Applied
  Physics 70~(2) (1991) 988--992.
\newblock \href {https://doi.org/10.1063/1.349611}
  {\path{doi:10.1063/1.349611}}.

\bibitem{sinha_xps_1987}
S.~Sinha, S.~Badrinarayanan, A.~Sinha, An {XPS} study of hydrogen implanted
  zirconium, Journal of the Less Common Metals 134~(2) (1987) 229--236.
\newblock \href {https://doi.org/10.1016/0022-5088(87)90562-5}
  {\path{doi:10.1016/0022-5088(87)90562-5}}.

\bibitem{kumar_xps_1988}
L.~Kumar, D.~Sarma, S.~Krummacher, {XPS} study of the room temperature surface
  oxidation of zirconium and its binary alloys with tin, chromium and iron,
  Applied Surface Science 32~(3) (1988) 309--319.
\newblock \href {https://doi.org/10.1016/0169-4332(88)90016-5}
  {\path{doi:10.1016/0169-4332(88)90016-5}}.

\bibitem{sarma_xpes_1980}
D.~Sarma, C.~Rao, {XPES} studies of oxides of second- and third-row transition
  metals including rare earths, Journal of Electron Spectroscopy and Related
  Phenomena 20~(1) (1980) 25--45.
\newblock \href {https://doi.org/10.1016/0368-2048(80)85003-1}
  {\path{doi:10.1016/0368-2048(80)85003-1}}.

\bibitem{morgan_comments_2021}
D.~J. Morgan, Comments on the {XPS} {Analysis} of {Carbon} {Materials}, C 7~(3)
  (2021) 51.
\newblock \href {https://doi.org/10.3390/c7030051}
  {\path{doi:10.3390/c7030051}}.

\bibitem{hauser_hybrid_2019}
D.~Hauser, A.~Auer, J.~Kunze-Liebhäuser, S.~Schwarz, J.~Bernardi, S.~Penner,
  Hybrid synthesis of zirconium oxycarbide nanopowders with defined and
  controlled composition, RSC Advances 9~(6) (2019) 3151--3156.
\newblock \href {https://doi.org/10.1039/C8RA09584A}
  {\path{doi:10.1039/C8RA09584A}}.

\bibitem{gu_promotion_2019}
H.~Gu, J.~Ding, Q.~Zhong, Y.~Zeng, F.~Song, Promotion of surface oxygen
  vacancies on the light olefins synthesis from catalytic {CO$_{2}$}
  hydrogenation over {FeK}/{ZrO$_{2}$} catalysts, International Journal of
  Hydrogen Energy 44~(23) (2019) 11808--11816.
\newblock \href {https://doi.org/10.1016/j.ijhydene.2019.03.046}
  {\path{doi:10.1016/j.ijhydene.2019.03.046}}.

\bibitem{liu_exploration_2017}
S.~Liu, X.~Wu, J.~Tang, P.~Cui, X.~Jiang, C.~Chang, W.~Liu, Y.~Gao, M.~Li,
  D.~Weng, An exploration of soot oxidation over {CeO$_{2}$}-{ZrO$_{2}$}
  nanocubes: {Do} more surface oxygen vacancies benefit the reaction?,
  Catalysis Today 281 (2017) 454--459.
\newblock \href {https://doi.org/10.1016/j.cattod.2016.05.036}
  {\path{doi:10.1016/j.cattod.2016.05.036}}.

\bibitem{ul-hamid_effect_2020}
A.~Ul-Hamid, The effect of deposition conditions on the properties of
  {Zr}-carbide, {Zr}-nitride and {Zr}-carbonitride coatings – a review,
  Materials Advances 1~(5) (2020) 988--1011.
\newblock \href {https://doi.org/10.1039/D0MA00232A}
  {\path{doi:10.1039/D0MA00232A}}.

\bibitem{badrinarayanan_xps_1989}
S.~Badrinarayanan, S.~Sinha, A.~Mandale, {XPS} studies of nitrogen ion
  implanted zirconium and titanium, Journal of Electron Spectroscopy and
  Related Phenomena 49~(3) (1989) 303--309.
\newblock \href {https://doi.org/10.1016/0368-2048(89)85018-2}
  {\path{doi:10.1016/0368-2048(89)85018-2}}.

\bibitem{calderon_v_chemical_2015}
S.~Calderon~V, A.~Cavaleiro, S.~Carvalho, Chemical and structural
  characterization of {ZrCNAg} coatings: {XPS}, {XRD} and {Raman} spectroscopy,
  Applied Surface Science 346 (2015) 240--247.
\newblock \href {https://doi.org/10.1016/j.apsusc.2015.03.161}
  {\path{doi:10.1016/j.apsusc.2015.03.161}}.

\bibitem{fu_synthesis_2004}
B.~Fu, L.~Gao, Synthesis of {Nanocrystalline} {Zirconium} {Nitride} {Powders}
  by {Reduction}-{Nitridation} of {Zirconium} {Oxide}, Journal of the American
  Ceramic Society 87~(4) (2004) 696--698.
\newblock \href {https://doi.org/10.1111/j.1551-2916.2004.00696.x}
  {\path{doi:10.1111/j.1551-2916.2004.00696.x}}.

\bibitem{zhao_synthesis_2019}
S.~Zhao, J.~Ma, R.~Xu, X.~Lin, X.~Cheng, S.~Hao, X.~Zhao, C.~Deng, B.~Liu,
  Synthesis and {Characterization} of {Zirconium} {Nitride} {Nanopowders} by
  {Internal} {Gelation} and {Carbothermic} {Nitridation}, Scientific Reports
  9~(1) (2019) 19199.
\newblock \href {https://doi.org/10.1038/s41598-019-55450-x}
  {\path{doi:10.1038/s41598-019-55450-x}}.

\bibitem{david_tem_2013}
J.~David, G.~Trolliard, M.~Gendre, A.~Maître, {TEM} study of the reaction
  mechanisms involved in the carbothermal reduction of zirconia, Journal of the
  European Ceramic Society 33~(1) (2013) 165--179.
\newblock \href {https://doi.org/10.1016/j.jeurceramsoc.2012.07.024}
  {\path{doi:10.1016/j.jeurceramsoc.2012.07.024}}.

\bibitem{sun_study_2018}
X.~Sun, C.~Deng, J.~Ma, X.~Zhao, S.~Hao, Z.~Li, B.~Liu, The study of
  carbothermic reduction–sintering of {ZrO$_{2}$}–{ZrC}–{C} composite
  microspheres prepared by internal gelation, Journal of Materials Science
  53~(20) (2018) 14149--14159.
\newblock \href {https://doi.org/10.1007/s10853-018-2646-0}
  {\path{doi:10.1007/s10853-018-2646-0}}.

\bibitem{sondhi_theoretical_2013}
A.~Sondhi, C.~Morandi, R.~Reidy, T.~Scharf, Theoretical and experimental
  investigations on the mechanism of carbothermal reduction of zirconia,
  Ceramics International 39~(4) (2013) 4489--4497.
\newblock \href {https://doi.org/10.1016/j.ceramint.2012.11.043}
  {\path{doi:10.1016/j.ceramint.2012.11.043}}.

\bibitem{harrison_processing_2016}
R.~W. Harrison, W.~E. Lee, Processing and properties of {ZrC}, {ZrN} and {ZrCN}
  ceramics: a review, Advances in Applied Ceramics 115~(5) (2016) 294--307.
\newblock \href {https://doi.org/10.1179/1743676115Y.0000000061}
  {\path{doi:10.1179/1743676115Y.0000000061}}.

\bibitem{parkison_deconvolution_2016}
A.~J. Parkison, A.~T. Nelson, Deconvolution of {Mass} {Gain} and {Mass} {Loss}
  {Mechanisms} {During} {Carbothermic} {Reduction} to {Nitridation} of
  {Zirconia}, Journal of the American Ceramic Society 99~(5) (2016) 1525--1533.
\newblock \href {https://doi.org/10.1111/jace.14156}
  {\path{doi:10.1111/jace.14156}}.

\bibitem{Gotsch2016}
T.~Götsch, W.~Wallisch, M.~Stöger-Pollach, B.~Klötzer, S.~Penner, From
  zirconia to yttria: Sampling the ysz phase diagram using sputter-deposited
  thin films, AIP Advances 6 (2 2016).
\newblock \href {https://doi.org/10.1063/1.4942818}
  {\path{doi:10.1063/1.4942818}}.

\bibitem{Gurak2018}
M.~Gurak, Q.~Flamant, L.~Laversenne, D.~R. Clarke, On the yttrium tantalate –
  zirconia phase diagram, Journal of the European Ceramic Society 38 (2018)
  3317--3324.
\newblock \href {https://doi.org/10.1016/j.jeurceramsoc.2018.03.012}
  {\path{doi:10.1016/j.jeurceramsoc.2018.03.012}}.

\bibitem{Block1985}
S.~Block, J.~A. H.~D. Jornada, G.~J. Piermarini, Pressure-temperature phase
  diagram of zirconia, Journal of the American Ceramic Society (1985) 497--99.

\bibitem{Jin2005}
X.~J. Jin, Martensitic transformation in zirconia containing ceramics and its
  applications, Current Opinion in Solid State and Materials Science 9 (2005)
  313--318.
\newblock \href {https://doi.org/10.1016/j.cossms.2006.02.012}
  {\path{doi:10.1016/j.cossms.2006.02.012}}.

\bibitem{lerch_nitridation_2005}
M.~Lerch, Nitridation of {Zirconia}, Journal of the American Ceramic Society
  79~(10) (2005) 2641--2644.
\newblock \href {https://doi.org/10.1111/j.1151-2916.1996.tb09028.x}
  {\path{doi:10.1111/j.1151-2916.1996.tb09028.x}}.

\bibitem{cheng_nitrogen-containing_1991}
Y.~Cheng, D.~P. Thompson, Nitrogen-{Containing} {Tetragonal} {Zirconia},
  Journal of the American Ceramic Society 74~(5) (1991) 1135--1138.
\newblock \href {https://doi.org/10.1111/j.1151-2916.1991.tb04355.x}
  {\path{doi:10.1111/j.1151-2916.1991.tb04355.x}}.

\bibitem{cheng_role_1993}
Y.-B. Cheng, D.~P. Thompson, Role of {Anion} {Vacancies} in
  {Nitrogen}-{Stabilized} {Zirconia}, Journal of the American Ceramic Society
  76~(3) (1993) 683--688.
\newblock \href {https://doi.org/10.1111/j.1151-2916.1993.tb03660.x}
  {\path{doi:10.1111/j.1151-2916.1993.tb03660.x}}.

\bibitem{humphrey}
F.~Humphreys, M.~Hatherly, Recrystallization and Related Annealing Phenomena,
  Elsevier, Oxford, 2004.

\end{thebibliography}


\begin{thebibliography}{1}
\expandafter\ifx\csname url\endcsname\relax
  \def\url#1{\texttt{#1}}\fi
\expandafter\ifx\csname urlprefix\endcsname\relax\def\urlprefix{URL }\fi
\expandafter\ifx\csname href\endcsname\relax
  \def\href#1#2{#2} \def\path#1{#1}\fi

\bibitem{Nakayama2017}
H.~Nakayama, K.~Ozaki, T.~Nabeta, Y.~Nakajima, Composition dependence of
  lattice parameter, thermal and electrical properties in zrcx compounds,
  Materials Transactions 58 (2017) 852--856.
\newblock \href {https://doi.org/10.2320/matertrans.M2016283}
  {\path{doi:10.2320/matertrans.M2016283}}.

\bibitem{Teeparthi2018}
S.~R. Teeparthi, E.~W. Awin, R.~Kumar, Dominating role of crystal structure
  over defect chemistry in black and white zirconia on visible light
  photocatalytic activity, Scientific Reports 8 (12 2018).
\newblock \href {https://doi.org/10.1038/s41598-018-23648-0}
  {\path{doi:10.1038/s41598-018-23648-0}}.

\bibitem{Gasparrini2018}
C.~Gasparrini, R.~J. Chater, D.~Horlait, L.~Vandeperre, W.~E. Lee, Zirconium
  carbide oxidation: Kinetics and oxygen diffusion through the intermediate
  layer, Journal of the American Ceramic Society 101 (2018) 2638--2652.
\newblock \href {https://doi.org/10.1111/jace.15479}
  {\path{doi:10.1111/jace.15479}}.

\bibitem{Sinhamahapatra2016}
A.~Sinhamahapatra, J.~P. Jeon, J.~Kang, B.~Han, J.~S. Yu, Oxygen-deficient
  zirconia (zro2-x): A new material for solar light absorption, Scientific
  Reports 6 (6 2016).
\newblock \href {https://doi.org/10.1038/srep27218}
  {\path{doi:10.1038/srep27218}}.

\end{thebibliography}





\end{document}


\renewcommand{\figurename}{Fig.}
\renewcommand{\thefigure}{S\arabic{figure}}

\setcounter{table}{0}
\renewcommand{\tablename}{Table}
\renewcommand{\thetable}{S\arabic{table}}

\maketitle

\section{Supplementary Note I: Material System}
CHN and ICP-MS analyses were used to determine carbon and zirconium content. CHN analysis was performed using a CE 440 Elemental Analyzer (Exeter Analytical Inc.; North Chelmsford, Massachusetts). ICP-MS analysis was performed with a NexION 350D (PerkinElmer Inc.; Waltham, Massachusetts). Carbon and zirconium atomic percentages were found to be approximately 39\% and 61\% respectively. To further verify the composition, x-ray diffraction (XRD) profiles (Cu K-$\alpha_{1}$ source; D8 Advance; Bruker Corperation, Billerica, Massachusetts) were used to compute equilibrium lattice parameter, from which the correlation proposed by Nakayama et al. \cite{Nakayama2017} was used to estimate carbon content. Lattice parameter was found to be 4.6788 \r{A}, corresponding to a carbon content of 41.2 at.\%. Bulk density of sample specimens was found to be 5.66 g/$cm^{3}$. Given the determined composition and density of the materials, samples are inferred to have a total porosity of approximately 12\%.

\section{Supplementary Note II: Characterization Methods}
Samples which retained a high degree of mechanical integrity in their oxide formation were then sectioned and polished for scanning electron microscopy (SEM, Quanta FEG 450 ESEM, FEI Company) and x-ray photolectron spectroscopy (XPS, Kratos Axis ULTRA, Kratos Analytical). SEM was performed in low-vacuum mode with a water dispersion at 1 Torr, with an accelerating voltage of 10 kV. XPS high resolution scans were performed using a monochromatic Al K$\alpha$ source (1486.6 eV) with a pass energy of 40 eV and a step size of 0.1 eV. The anaylsis chamber was maintained at ~1.7 × 10$^{-7}$ Torr. Spectra were averaged over ten scans and charge corrected to the adventitious carbon peak at 284.8 eV. A portion of select oxide formations was also removed and processed in an agate mortar and pestle for x-ray powder diffraction (Cu K-$\alpha_{1}$ source; D8 Advance; Bruker Corperation).

\section{Supplementary Note III: Test Matrix}
Specimen ID, isothermal oxidation temperature, exposure duration, initial mass (m$_0$), initial dimensions, and final mass (m$_f$) are described in Table \ref{test_matrix}.

\begin{sidewaystable*}[h]
\centering
\begin{tabular}{c c c c c c}
\toprule
Specimen ID & Temperature (°C) & Exposure Duration (min) & m$_0$ (g) & Initial Dimensions (mm) & m$_f$ (g) \\ 
\midrule \hline
    ZrC-001 & 1000 & 10 & 1.90 & 10.58 x 6.33 x 5.16 & 1.91 \\
    ZrC-002 & 1000 & 20 & 2.35 & 11.92 x 6.30 x 5.44 & 2.42 \\
    ZrC-003 & 1000 & 30 & 2.41 & 12.36 x 6.30 x 5.40 & 2.54 \\
    ZrC-004 & 1000 & 40 & 3.03 & 12.96 x 6.30 x 6.52 & 3.20 \\
    ZrC-005 & 1200 & 10 & 2.40 & 12.23 x 6.35 x 5.49 & 2.43 \\
    ZrC-006 & 1200 & 20 & 2.37 & 12.09 x 6.31 x 5.50 & 2.45 \\
    ZrC-007 & 1200 & 30 & 2.59 & 12.44 x 6.35 x 5.85 & 2.77 \\
    ZrC-008 & 1200 & 40 & 2.45 & 12.05 x 6.32 x 5.80 & 2.63 \\
    ZrC-009 & 1400 & 10 & 2.21 & 10.71 x 6.34 x 5.73 & 2.27 \\
    ZrC-010 & 1400 & 20 & 2.67 & 13.12 x 6.51 x 5.84 & 2.77 \\
    ZrC-011 & 1400 & 30 & 2.50 & 12.16 x 6.32 x 5.77 & 2.65 \\
    ZrC-012 & 1400 & 40 & 2.62 & 12.71 x 6.32 x 5.80 & 2.79 \\
    ZrC-013 & 1500 & 10 & 2.64 & 12.45 x 6.36 x 5.88 & 2.68 \\
    ZrC-014 & 1500 & 20 & 2.97 & 12.52 x 6.35 x 6.61 & 3.05 \\
    ZrC-015 & 1500 & 30 & 2.02 & 10.98 x 6.35 x 5.33 & 2.11 \\
    ZrC-016 & 1500 & 40 & 2.60 & 12.65 x 6.32 x 5.78 & 2.74 \\
    ZrC-017 & 1600 & 10 & 2.72 & 11.97 x 6.33 x 6.51 & 2.80 \\
    ZrC-018 & 1600 & 20 & 2.63 & 11.83 x 6.33 x 6.37 & 2.75 \\
    ZrC-019 & 1600 & 30 & 2.51 & 11.66 x 6.34 x 6.12 & 2.66 \\
    ZrC-020 & 1600 & 40 & 2.93 & 12.30 x 6.31 x 6.63 & 3.09 \\
\midrule \hline
\end{tabular}
\caption{Test Matrix}
\label{test_matrix}
\end{sidewaystable*}

\section{Supplementary Note IV: X-ray Photoelectron Spectroscopy}
\begin{figure}
    \centering
    \includegraphics[width=0.7\textwidth]{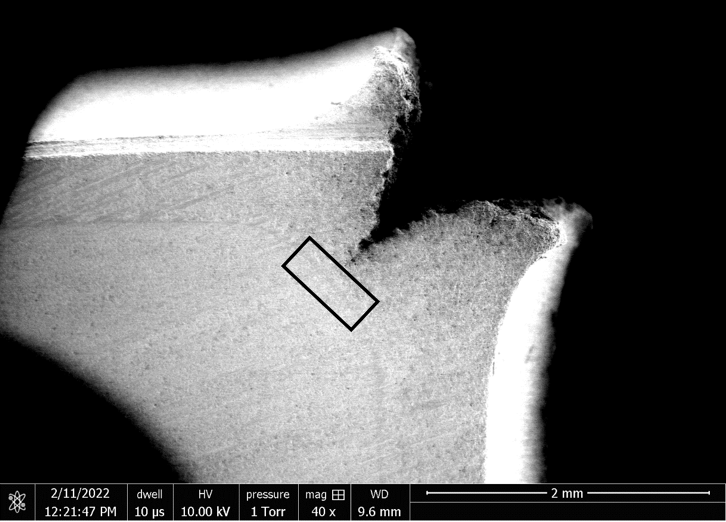}
    \caption{To scale visualization of x-ray photoelectron spectroscopy incident area on the cross section of a material specimen.}
    \label{XPS_spot}
\end{figure}

For specimens oxidized at sufficiently high temperatures (1200 $^{\circ}$C and above) such that the specimen was able to be sectioned and polished, the probe area (0.3 by 0.7mm) was focused on an identical location on the corner of a given cross section, encompassing the oxide, intermediate, and substrate in order to get a holistic view of the reaction front as seen in Fig. \ref{XPS_spot}. 

\section{Supplementary Note V: Mechanical Integrity}
Referencing Figure 1b in the text: at 1000 $^{\circ}$C, the oxide layer lost all mechanical integrity and was primarily powder upon removal from the furnace. The image shown is the collected scale powder stored in a bag and is off-white in color (with white being typical of a pure monoclinic zirconia, whereas a much darker color is characteristic of a tetragonal phase \cite{Teeparthi2018}). This is attributed to m-ZrO$_{2}$ formation during oxidation of the specimen, implying no t- to m-ZrO$_{2}$ transition during cooling and therefore no transformation toughening present in the oxide formation, at least in part, leading to the loss of structural integrity. At 1200 $^{\circ}$C we observe a relatively stable oxide formation, yet with some powder present and a similar off-white color as seen in the 1000 $^{\circ}$C specimen. Here, a polymorphic oxide formation is likely to have formed during oxidation, meaning a portion of the oxide underwent phase transformation toughening upon cooling, while other formations did not. Finally, at all temperatures above 1400 $^{\circ}$C, we observe very similar oxides, which are highly stable and grey in color –  attributed to highly dominant t-ZrO$_{2}$ nucleation during oxidation \cite{Gasparrini2018} \cite{Sinhamahapatra2016}, with the vast majority (if not all) of the oxide volume undergoing a Martensitic transformation upon cooling. 

 \bibliographystyle{elsarticle-num} 
 \bibliography{refs}